\documentclass[12pt]{article}
\usepackage{amssymb,graphicx}

\def\np{\vspace{12pt} \noindent}

\begin{document}

\begin{center}
{\Large \bf {Old Galaxies in the Young Universe}}
\end{center}
\begin{center}
{\bf {A. Cimatti$^1$, E. Daddi$^2$, A. Renzini$^2$, P. Cassata$^3$,
E. Vanzella$^{3}$, L. Pozzetti$^4$, S. Cristiani$^5$, A. Fontana$^6$, 
G. Rodighiero$^3$, M. Mignoli$^4$, G. Zamorani$^4$
}}\\

\medskip
\noindent
\begin{small}
$^1$ INAF - Osservatorio Astrofisico di Arcetri, Largo E. Fermi 5,
I-50125, Firenze, Italy \\ 
\np
$^2$ European Southern Observatory, Karl-Schwarzschild-Str. 2, D-85748,
Garching, Germany \\ 
\np
$^3$ Dipartimento di Astronomia, Universit\`a di Padova, Vicolo
dell'Osservatorio, 2, I-35122 Padova, Italy \\
\np
$^4$ INAF - Osservatorio Astronomico di Bologna, via Ranzani 1, I-40127,
Bologna, Italy\\ 
\np
$^5$ INAF - Osservatorio Astronomico di Trieste, Via Tiepolo 11, I-34131
Trieste, Italy\\ 
\np
$^6$ INAF - Osservatorio Astronomico di Roma, via dell'Osservatorio 2, 
Monteporzio, Italy
\end{small}
\end{center}

{\np \bf 
More than half of all stars in the local Universe are found in massive 
spheroidal galaxies$^{1}$, which are characterized by old stellar 
populations$^{2,3}$ with little or no current star formation. In 
present models, such galaxies appear rather late as the culmination of a 
hierarchical merging process, in which larger galaxies are assembled 
through mergers of smaller precursor galaxies. But observations have not yet 
established how, or even when, the massive spheroidals formed$^{2,3}$, nor
if their seemingly sudden appearance when the Universe was about half 
its present age (at redshift $z \approx 1$) results from a real 
evolutionary effect (such as a peak of mergers) or from the observational 
difficulty of identifying them at earlier epochs. 
Here we report the spectroscopic and morphological identification of 
four old, fully assembled, massive ($>10^{11}$ solar masses) spheroidal 
galaxies at $1.6<z<1.9$, the most distant such objects currently known. 
The existence of such systems when the Universe was only one-quarter of 
its present age, shows that the build-up of massive early-type galaxies 
was much faster in the early Universe than has been expected from 
theoretical simulations$^{4}$.
}

\np

In the $\Lambda$CDM scenario$^5$, galaxies are thought to build-up 
their present-day mass through a continuous assembly driven by the
hierarchical merging of dark matter halos, with the most massive
galaxies being the last to form. However, the formation and evolution 
of massive spheroidal early-type galaxies is still an open question.

Recent results indicate that early-type galaxies are found up to
$z\sim1$ with a number density comparable to that of local luminous
E/S0 galaxies$^{6,7}$, suggesting a slow evolution of their
stellar mass density from $z\sim1$ to the present epoch. The
critical question is whether these galaxies do exist in substantial 
number$^{8,9}$ at earlier epochs, or if they were assembled later$^{10,11}$ 
as favored by most renditions of the hierarchical galaxy formation 
scenario$^{4}$. The problem is complicated also by the difficulty of 
identifying such galaxies due to their faintness and, for $z>1.3$, 
the lack of strong spectral features in optical spectra, placing them 
among the most difficult targets even for the largest optical telescopes. 
For example, while star-forming galaxies are now routinely found up to 
$z\sim6.6$$^{12}$, the most distant spectroscopically confirmed old 
spheroid is still a radio--selected object at $z=1.552$ discovered 
almost a decade ago$^{13,14}$. 

One way of addressing the critical question of massive galaxy formation 
is to search for the farthest and oldest galaxies with masses comparable 
to the most massive galaxies in the present-day universe ($10^{11-12}$ 
M$_{\odot}$), and to use them as the ``fossil'' tracers of the most 
remote events of galaxy formation. As the rest-frame optical -- near-infrared 
luminosity traces the galaxy mass$^{15}$, the $K_s$-band 
($\lambda \sim 2.2\,\mu$m in the observer frame) allows a fair selection 
of galaxies according to their mass up to $z\sim 2$. 

Following this approach, we recently conducted the K20 survey$^{16}$ with 
the Very Large Telescope (VLT) of the European Southern Observatory 
(ESO). Deep optical spectroscopy was obtained for a sample of 546 
objects with $K_s<20$ (Vega photometric scale) and extracted from an 
area of 52 arcmin$^2$, including 32 arcmin$^2$ within the GOODS--South
field $^{17}$ (hereafter the GOODS/K20 field). The spectroscopic 
redshift ($z_{spec}$) completeness of the K20 survey is 92\%, while the 
available multi-band photometry ($BVRIzJHK_s$) allowed us to derive the
spectral energy distribution (SED) and photometric redshift ($z_{phot}$)
of each galaxy. The K20 survey spectroscopy was complemented with the 
ESO/GOODS public spectroscopy (Supplementary Table 1). 

The available spectra within the GOODS/K20 field were then used to search 
for old, massive galaxies at $z>1.5$. We spectroscopically identified 
four galaxies 
with $18 \lesssim K_s \lesssim 19$ and $1.6 \lesssim z_{spec} \lesssim
1.9$ which have rest-frame mid-UV spectra with shapes
and continuum breaks compatible with being dominated by old
stars and $R-K_s \gtrsim 6$ (the colour expected at $z>1.5$ for
old passively evolving galaxies due to the combination of old stellar
populations and k-correction effects$^{9}$). The Supplementary Table 
1 lists the main
galaxy information. The spectrum of each individual object allows a 
fairly precise determination of the redshift based on absorption 
features and on the overall spectral shape (Fig. 1).

The co-added average spectrum of the four galaxies (Fig. 2--3) shows a
near-UV continuum shape, breaks and absorption lines that are intermediate 
between those of a F2 V and a F5 V star$^{18}$, and typical of about 1-2 
Gyr old synthetic stellar populations$^{19,20}$. It is also very 
similar to the average spectrum of $z\sim1$ old Extremely 
Red Objects$^7$ (EROs), and slightly bluer than that of the $z\sim0.5$ 
SDSS red luminous galaxies$^{21}$ and of the $z=1.55$ old galaxy LBDS 
53w091$^{13}$. However, it is different in shape and slope from 
the average spectrum of $z\sim1$ dusty star-forming EROs$^7$. 

The multi-band photometric SED of each galaxy was successfully fitted 
without the need for dust extinction, and using a library of simple 
stellar population (SSP) models$^{19}$ with a wide range of ages,
$Z=Z_{\odot}$ and Salpeter IMF. This procedure yielded best-fitting 
ages of 1.0-1.7 Gyr, the mass-to-light ratios and hence the stellar mass 
of each galaxy, which results in the range of 1--3$\times 10^{11}$ 
$h_{70}^{-2}$ M$_{\odot}$. $H_0=70$ km s$^{-1}$ Mpc$^{-1}$ 
(with $h_{70} \equiv H_0/70$), $\Omega_{\rm m}=0.3$ and $\Omega_{\Lambda}=0.7$ 
are adopted.

In addition to spectroscopy, the nature of these galaxies was
investigated with the fundamental complement of {\sl Hubble Space Telescope}+
ACS ({\sl Advanced Camera for Surveys}) imaging from the GOODS public
{\sl Treasury Program}$^{17}$. The analysis of the ACS high-resolution 
images reveals that the surface brightness distribution of these galaxies 
is typical of elliptical/early-type galaxies (Fig. 4).

Besides pushing to $z\sim1.9$ the identification of the highest
redshift elliptical galaxy, these objects are very relevant to understand
the evolution of galaxies in general for three main reasons: their old age, 
their high mass, and their substantial number density.

Indeed, an average age of about 1-2 Gyr ($Z=Z_{\odot}$) at $<\! z\!>\sim 
1.7$ implies that the onset of the star formation occurred not later 
than at $z\sim 2.5-3.4$ ($z\sim 2-2.5$ for $Z=2.5Z_{\odot}$). 
These are strict lower limits 
because they follow from assuming instantaneous bursts, whereas a more 
realistic, prolonged star formation activity would push the bulk of 
their star formation to an earlier cosmic epoch. As an illustrative
example, the photometric SED of ID 646 ($z=1.903$) can be reproduced 
(without dust) with either a $\sim$1 Gyr old instantaneous burst 
occurred at $z \sim 2.7$, or with a $\sim$2 Gyr old stellar population 
with a star formation rate declining with $exp(-t/ \tau)$ ($\tau=0.3$ Gyr). 
In the latter case, the star formation onset would be pushed to $z \sim 4$
and half of the stars would be formed at $z \sim 3.6$.
In addition, with stellar masses $M_*>10^{11} h_{70}^{-2} M_\odot$, 
these systems 
would rank among the most massive galaxies in the present-day universe, 
suggesting that they were fully assembled already at this early epoch.

Finally, their number density is considerably high. Within the comoving 
volume relative to 32 arcmin$^2$ and $1.5<z<1.9$
(40,000 $h_{70}^{-3}$ Mpc$^3$), the comoving density of such galaxies
is about $10^{-4}$ $h_{70}^{3}$ Mpc$^{-3}$, corresponding to a stellar mass
density of about $2 \times 10^{7}$ $h_{70}$ M$_{\odot}$Mpc$^{-3}$, 
i.e. about 10\% of the local ($z=0$) value$^{22}$ for masses greater
than $10^{11}$ M$_{\odot}$. This mass density is comparable to 
that of star-forming $M_*>10^{11}M_\odot$ galaxies at $z\sim 2$ $^{23}$, 
suggesting that while the most massive galaxies in the local universe 
are now old objects with no or weak star formation, by $z\sim 2$ passive 
and active star-forming massive galaxies coexist in nearly equal number. 

Although more successful than previous models, the most recent realizations 
of semi-analytic hierarchical merging simulations still severely
underpredict the density of such old galaxies: just one old galaxy with 
$K_s<20$, $R-K_s>6$, and $z>1.5$ is present in the mock 
catalog$^{4}$ for the whole five times wider GOODS/CDFS area. 

As expected for early-type galaxies$^{9,24}$, the three galaxies at 
$z\sim 1.61$ may trace the underlying large scale structure.
In this case, our estimated number density may be somewhat biased toward
a high value. On the other hand, the number of such galaxies in our
sample is likely to be a lower limit due to the spectroscopic redshift 
incompleteness. There are indeed up to three more candidate old galaxies in 
the GOODS/K20 sample with $18.5 \lesssim K_s \lesssim 19.5$, $1.5 \lesssim 
z_{phot} \lesssim 2.0$, $5.6 \lesssim R-K_s \lesssim 6.8$ and compact HST 
morphology. Thus, in the GOODS/K20 sample the fraction of old galaxies 
among the whole $z>1.5$ galaxy population is 15$\pm$8\% (spectroscopic 
redshifts only), or up to 25$\pm$11\% if also all the 3 additional 
candidates are counted.

It is generally thought that the so-called ``redshift desert''
(i.e. around $1.4<z<2.5$) represents the cosmic epoch when most star formation
activity and galaxy mass assembly took place$^{25}$. Our results show
that, in addition to actively star forming galaxies$^{26}$, also a 
substantial number of ``fossil'' systems already populate this redshift 
range, and hence remain undetected in surveys biased towards star-forming 
systems. The luminous star-forming galaxies found at $z>2$ in 
sub-mm$^{27}$ and near-infrared$^{23,28}$ surveys may represent the 
progenitors of these old and massive systems. 

\np
1. Fukugita, M., Hogan, C.J., Peebles, P.J.E. The Cosmic Baryon Budget.
Astrophys. J. 503, 518-530 (1998).\\
2. Renzini, A. Origin of Bulges. In ``The formation of galactic
bulges'', ed. C.M. Carollo, H.C. Ferguson, R.F.G. Wyse,  Cambridge
University Press, p.9-26 (1999).\\
3. Peebles, P.J.E. When did the Large Elliptical Galaxies Form?
In ``A New Era in Cosmology'', ASP Conference Proceedings, Vol. 283.
ed. N. Metcalfe and T. Shanks, Astronomical Society of the Pacific,
2002., p.351-361 (2002).\\
4. Somerville, R.S. et al. The Redshift Distribution of
Near-Infrared-Selected Galaxies in the Great Observatories Origins Deep
Survey as a Test of Galaxy Formation Scenarios. Astrophys. J., 600, L135-139
(2004).\\
5. Freedman, W.L. \& Turner, M.S. Colloquium: Measuring and
understanding the universe. Reviews of Modern Physics, 75, 
1433-1447 (2003).\\
6. Im, M. et al. The DEEP Groth Strip Survey. X. Number Density and
Luminosity Function of Field E/S0 Galaxies at $z<1$. Astrophys. J. 571,
136-171 (2002).\\
7. Cimatti, A. et al. The K20 survey. I. Disentangling old and dusty
star-forming galaxies in the ERO population. Astron. Astrophys. 381, L68-73
(2002).\\
8. Benitez, N. et al. Detection of Evolved High-Redshift Galaxies in
Deep NICMOS/VLT Images. Astrophys. J. 515, L65-69 (1999).\\
9. Daddi, E. et al. Detection of strong clustering of extremely red
objects: implications for the density of $z>1$ ellipticals. Astron.
Astrophys. 361, 535-549 (2000).\\
10. Zepf, S.E. Formation of elliptical galaxies at moderate redshifts.
Nature 390, 377-380 (1997).\\
11. Rodighiero G., Franceschini A., Fasano G. Deep Hubble Space Telescope
imaging surveys and the formation of spheroidal galaxies. Mon. Not. R.
Astron. Soc. 324, 491-497 (2001).\\
12. Taniguchi, Y. et al. Lyman$\alpha$ Emitters beyond Redshift 5: The
Dawn of Galaxy Formation. Journal of the Korean Astronomical Society 36,
no.3, 123-144 (2003).\\
13. Dunlop, J.S. et al. A 3.5-Gyr-old galaxy at redshift 1.55. Nature,
381, 581-584 (1996).\\
14. Spinrad, H., Dey, A., Stern, D., Dunlop, J., Peacock, J., Jimenez, 
R., Windhorst, R. LBDS 53W091: an Old, Red Galaxy at z=1.552. Astrophys.
J., 484, 581-601 (1997).\\
15. Gavazzi, G., Pierini, D., Boselli, A., The phenomenology of disk
galaxies. Astron. Astrophys. 312, 397-408 (1996).\\
16. Cimatti, A. et al. The K20 survey. III. Photometric and
spectroscopic properties of the sample. Astron. Astrophys. 392, 395-406
(2002).\\
17. Giavalisco, M. et al., The Great Observatories Origins Deep Survey:
Initial Results from Optical and Near-Infrared Imaging. Astrophys. J. 600,
L93-98 (2004).\\
18. Pickles, A.J. A Stellar Spectral Flux Library: 1150-25000~\AA.
PASP 110, 863-878 (1998).\\
19. Bruzual, G. \& Charlot, S. Stellar population synthesis at the
resolution of 2003. Mon. Not. R. Astron. Soc. 344, 1000-1028 (2003).\\
20. Jimenez, R. et al. Synthetic stellar populations: single stellar
populations, stellar interior models and primordial proto-galaxies,
Mon. Not. R. Astron. Soc. 349, 240-254 (2004).\\
21. Eisenstein, D.J. et al. Average Spectra of Massive Galaxies in the
Sloan Digital Sky Survey. Astrophys. J. 585, 694-713 (2003).\\ 
22. Cole, S. et al. The 2dF galaxy redshift survey: near-infrared galaxy
luminosity functions. Mon. Not. R. Astron. Soc., 326, 255-273 (2001).\\
23. Daddi, E. et al., Near-Infrared Bright Galaxies at $z\sim2$. Entering
the Spheroid Formation Epoch ? Astrophys. J., 600, L127-131 (2004).\\
24. Davis, M., Geller, M.J. Galaxy Correlations as a Function of
Morphological Type. Astrophys. J., 208, 13-19 (1976).\\
25. Dickinson, M., Papovich, C., Ferguson, H.C., Budavari, T. The
Evolution of the Global Stellar Mass Density at $0<z<3$. Astrophys. J.,
587, 25-40 (2003).\\
26. Steidel, C.C. et al. A Survey of Star-Forming Galaxies in the
z=1.4-2.5 `Redshift Desert': Overview. Astrophys. J. 604, 534-550 (2004).\\
27. Genzel, R., Baker, A.J., Tacconi, L.J., Lutz, D., Cox, P.; Guilloteau, 
S., Omont, A. Spatially Resolved Millimeter Interferometry of SMM
J02399-0136: A Very Massive Galaxy at $z=2.8$. Astrophys. J. 584, 633-642
(2003).\\
28. Franx, M. et al. A Significant Population of Red, Near-Infrared-selected 
High-Redshift Galaxies. Astrophys. J., 587, L79-L83 (2003).\\
29. Pignatelli, E. \& Fasano, G. GASPHOT: A Tool for Automated Surface
Photometry of Galaxies. Astrophys. Sp. Sci. 269, 657-658 (1999).\\
30. Peng, C.Y., Ho, L.C., Impey, C.D., Rix, H.-W., Detailed Structural
Decomposition of Galaxy Images. Astron. J., 124, 266-293 (2002). \\ 

\np

Correspondence and requests for material should be sent to Andrea
Cimatti (cimatti@arcetri.astro.it).

\np

This work is based on observations made at the European Southern
Observatory, Paranal, Chile, and with the NASA/ESA Hubble Space 
Telescope obtained at the Space Telescope Science Institute, which 
is operated by the Association of Universities for Research in 
Astronomy (AURA). We thank Rachel Somerville for information 
on the GOODS/CDFS mock catalog. We are grateful to the GOODS Team
for obtaining and releasing the HST and FORS2 data.

\pagebreak

{\bf Figure 1}

{\it The individual and average spectra of the detected galaxies.}
From bottom to top: the individual spectra smoothed to a
16~\AA~ boxcar (26~\AA~ for ID 237) and the average spectrum of the 
four old galaxies ($z_{average}=1.68$). The red line 
is the spectrum of the old galaxy LBDS 53w091 ($z=1.55$) used to 
search for spectra with a similar continuum shape. Weak 
features in individual spectra (e.g. MgII$\lambda$2800 and the 2640~\AA~
continuum break, B2640) become clearly visible in the average spectrum. 
The object ID 235 has also a weak [OII]$\lambda$3727
emission (not shown here). The spectra were obtained with 
ESO VLT+FORS2, grisms 200I (R($1^{\prime\prime})\sim$400) (ID 237) 
and 300I (R($1^{\prime\prime})\sim$600) (IDs 235,270,646), 
1.0$^{\prime\prime}$ wide slit and 
$\lesssim 1^{\prime\prime}$ seeing conditions. The integrations times were 
3 hours for ID 237, 7.8 hours for IDs 235 and 270, and 15.8 hours for 
ID 646. For ID 646, the ESO/GOODS public spectrum was 
co-added to our K20 spectrum (see Supplementary Tab. 1). ``Dithering'' 
of the targets along 
the slits was applied to remove efficiently the CCD fringing pattern 
and the strong OH sky lines in the red. The data reduction was done 
with the IRAF software package (see$^{16}$). 
The spectrophotometric calibration of
all spectra was achieved and verified by observing several standard stars. 
The average spectrum, corresponding to 34.4 hours integration time,
was obtained by co-adding the individual spectra convolved to the same 
resolution, scaled to the same arbitrary flux (i.e. with each spectrum 
having the same weight in the co-addition), and assigning 
wavelength--dependent weights which take into account the noise in 
the individual spectra due to the OH emission sky lines.

\np

\begin{figure}[t]
\begin{center}
\includegraphics[width=15cm]{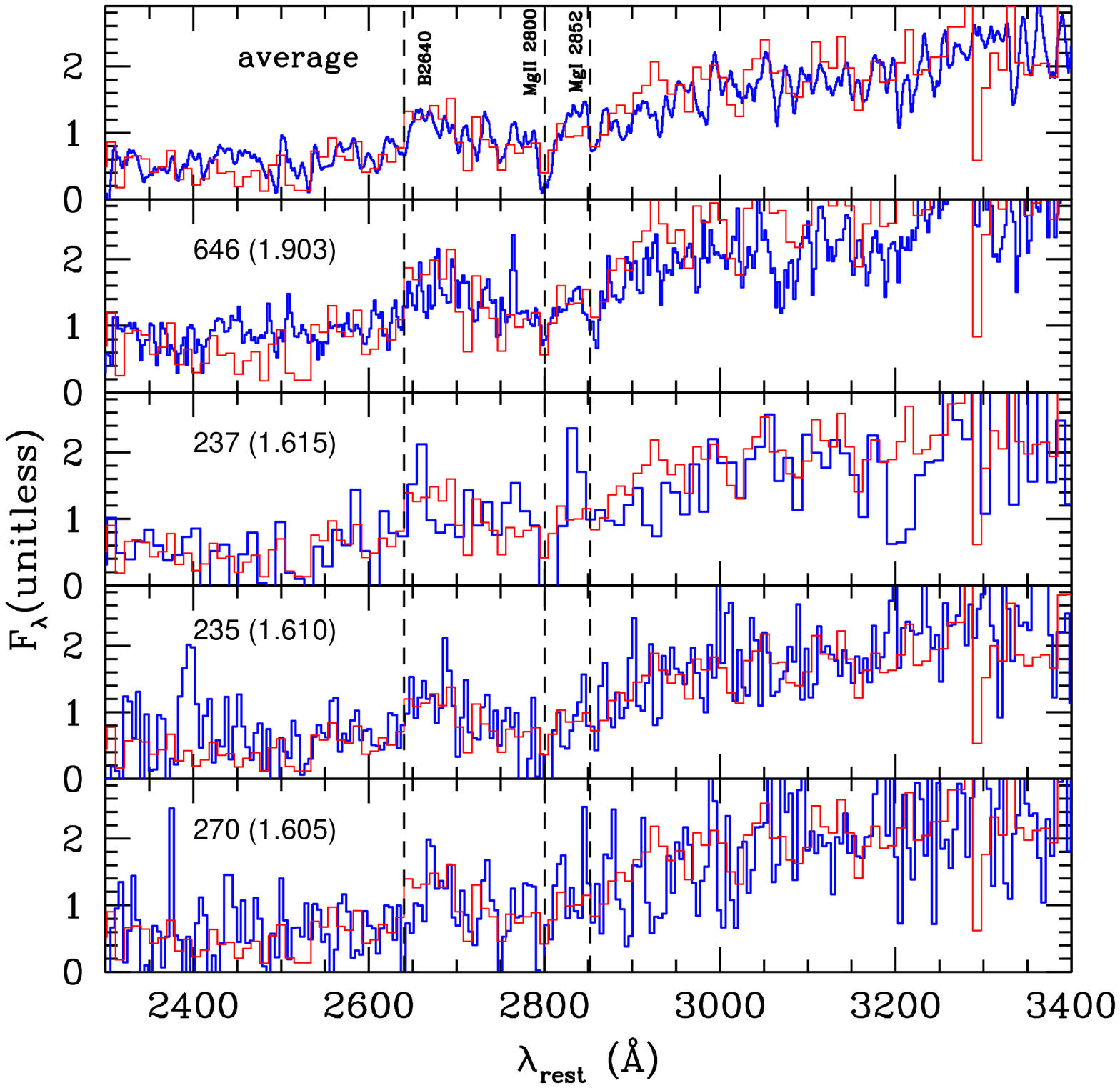}
\caption{}
\label{fig1}
\end{center}
\end{figure}

{\bf Figure 2}

{\it The detailed average spectrum of the detected galaxies.}
A zoom on the average spectrum (blue) compared with the synthetic 
spectrum$^{19}$ of a 1.1 Gyr old simple stellar population (SSP) 
with solar metallicity ($Z=Z_{\odot}$) and Salpeter IMF (red).
The observed average spectrum was compared to a library of synthetic
SSP template spectra$^{19,20}$ with a range of ages of 0.1-3.0 Gyr with a step 
of 0.1 Gyr, and with assumed metallicities $Z$=0.4$\times$, 1.0$\times$, and 
2.5$\times Z_{\odot}$. The best fit age for each set of synthetic 
templates was derived through a $\chi^2$ minimization over the rest-frame 
wavelength range 2300--3400~\AA. The rms as a function of wavelength 
used in the $\chi^2$ procedure was estimated from the average 
spectrum computing a running mean rms with a step of 1~\AA~ and a 
box size of 20~\AA, corresponding to about three times the resolution 
of the observed average spectrum. The median signal-to-noise ratio
is $\sim$20 per resolution element in the 2300--3400~\AA~ range. 
The wavelength ranges including the strongest real features (i.e. 
absorptions and continuum breaks) were not used in the estimate of 
the rms. The resulting reduced $\chi^2$
is of the order of unity for the best fit models. In the case of solar
metallicity, the ranges of ages acceptable at 95\% confidence level
are $1.0^{+0.5}_{-0.1}$ Gyr and $1.4^{+0.5}_{-0.4}$ Gyr for SSP models 
of$^{19}$ and$^{20}$ respectively (see also Fig. 3, top panel). 
Ages $\sim 50\%$ younger or older are also acceptable for $Z=2.5Z_{\odot}$ 
or $Z=0.4Z_{\odot}$ respectively. 
The 2640~\AA~ and 2900~\AA~ continuum break$^{13}$ amplitudes measured
on the average spectrum are B2640=1.8$\pm$0.1 and B2900=1.2$\pm$0.1. 
These values are consistent with the ones expected in SSP models$^{19-20}$ 
for ages around 1--1.5 Gyr and solar metallicity. For instance, the SSP 
model spectrum shown here has B2640=1.84 and B2900=1.27.  

\np

\begin{figure}[t]
\begin{center}
\includegraphics[width=15cm]{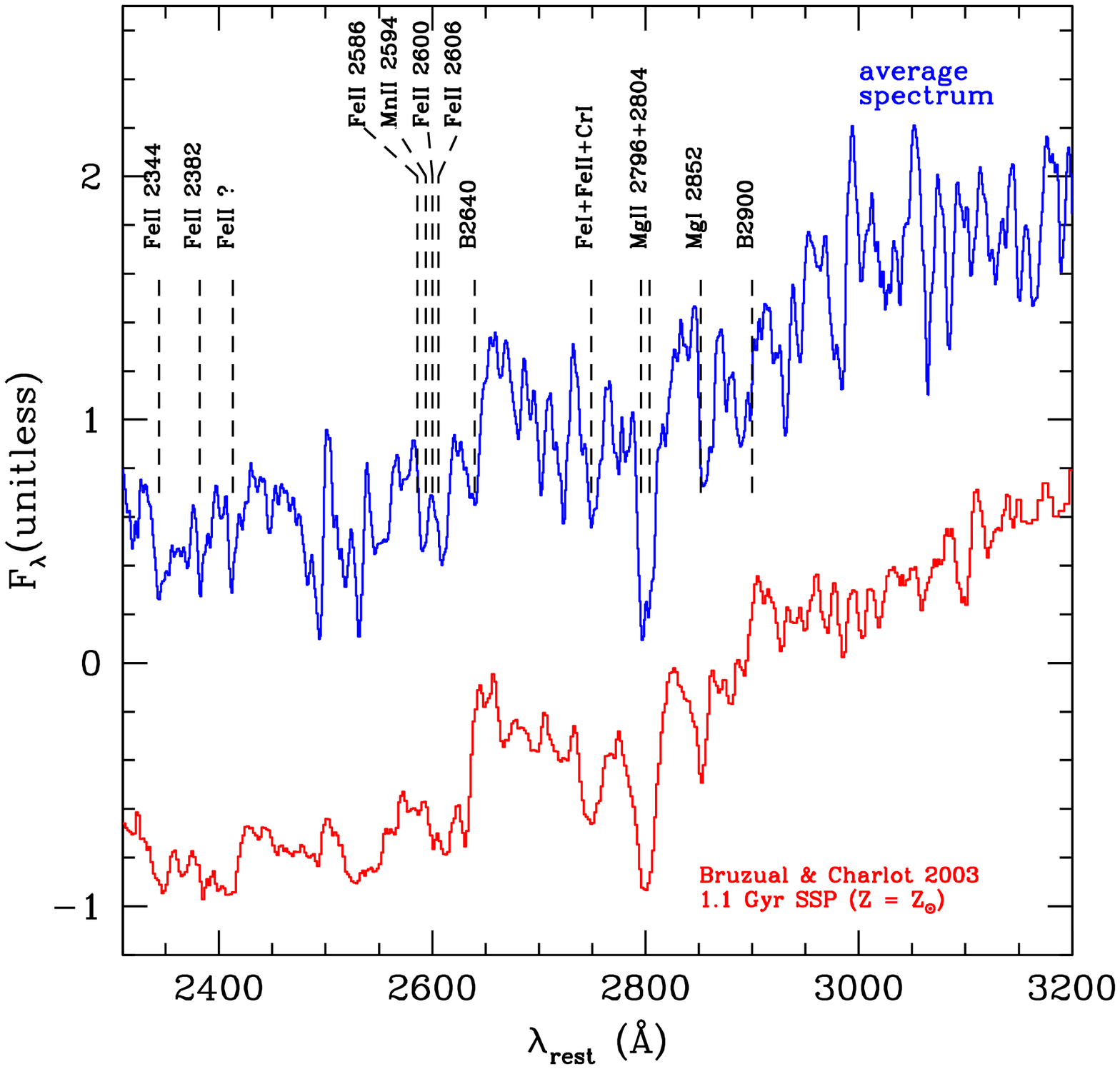}
\caption{}
\label{fig2}
\end{center}
\end{figure}

{\bf Figure 3}

{\it The comparison between the average spectrum and a set of
spectral templates.}
The average spectrum (blue) compared to a set of template
spectra. From bottom: F2 V (green) and F5 V (red) stellar spectra$^{18}$
with $Z=Z_{\odot}$, the composite spectrum (red) of 726 luminous red 
galaxies at $0.47<z<0.55$ selected from the SDSS$^{21}$ (available 
only for $\lambda>2600$~\AA), the average spectra of $z\sim1$ old
(red) and dusty star-forming (green) EROs$^7$, SSP synthetic 
spectra$^{19}$ ($Z=Z_{\odot}$, Salpeter IMF) with ages of 0.5 Gyr 
(magenta), 1.1 Gyr (green) and 3.0 Gyr (red).

\np

\begin{figure}[t]
\begin{center}
\includegraphics[width=15cm]{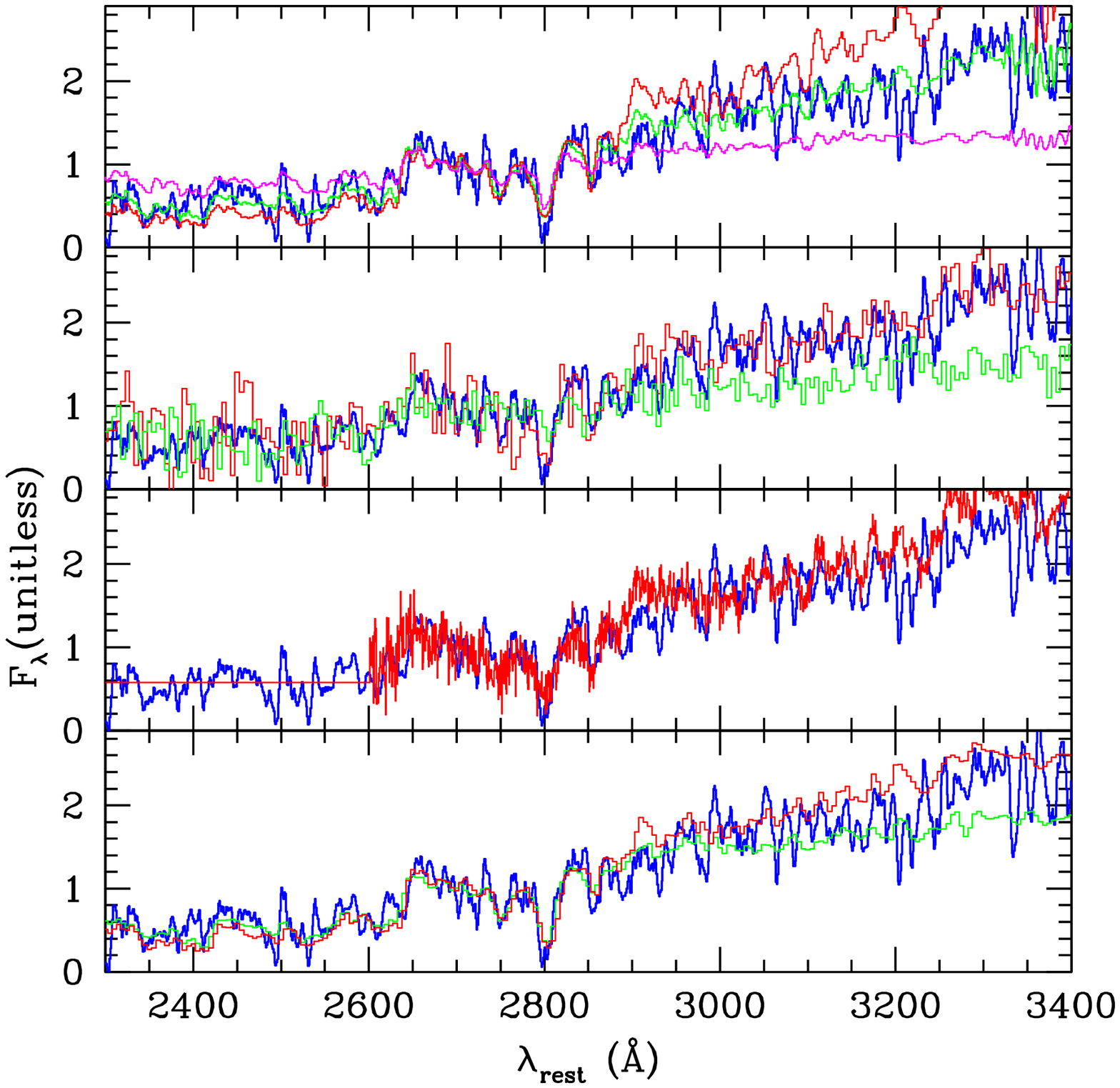}
\caption{}
\label{fig3}
\end{center}
\end{figure}

{\bf Figure 4}

{\it The morphological properties of the detected galaxies.}
Images of the four galaxies taken with the {\sl Hubble Space Telescope}
+ACS through the F850LP filter (from GOODS data$^{17}$)
which samples the rest-frame $\sim$3000-3500~\AA~ for $1.6<z<2$. 
The images are in logarithmic grey--scale and their size is 
$2^{\prime\prime} \times 2^{\prime\prime}$, corresponding to
$\sim 17 \times 17$ kpc for the average redshift $z=1.7$ and the 
adopted cosmology. 
At a visual inspection, the galaxies show rather compact morphologies
with most of the flux coming from the central regions. 
A fit of their surface brightness profiles was performed 
with a ``Sersic law'' ($\propto r^{1/n}$) 
convolved with the average point spread function extracted
from the stars in the ACS field and using the GASPHOT$^{29}$ and
GALFIT$^{30}$ software packages.  Objects ID 237 and ID 646 have
profiles with acceptable values of $n$ in the range of $4<n<6$, i.e.,
typical of elliptical galaxies, object ID 270 is better reproduced by
a flatter profile ($1<n<2$), whereas a more ambiguos result is found
for the object showing some evidence of irregularities in the
morphology (ID 235, $1<n<3$). These latter objects may be
bulge-dominated spirals but no bulge/disk decomposition was
attempted. Ground-based near-infrared images taken under
0.5$^{\prime\prime}$ seeing conditions with the ESO VLT+ISAAC 
through the $K_s$ filter (rest-frame $\sim$6000-8000~\AA) show
very compact morphologies, but no surface brightness fitting was done.  

\np

\begin{figure}[t]
\begin{center}
\includegraphics[width=15cm,angle=180]{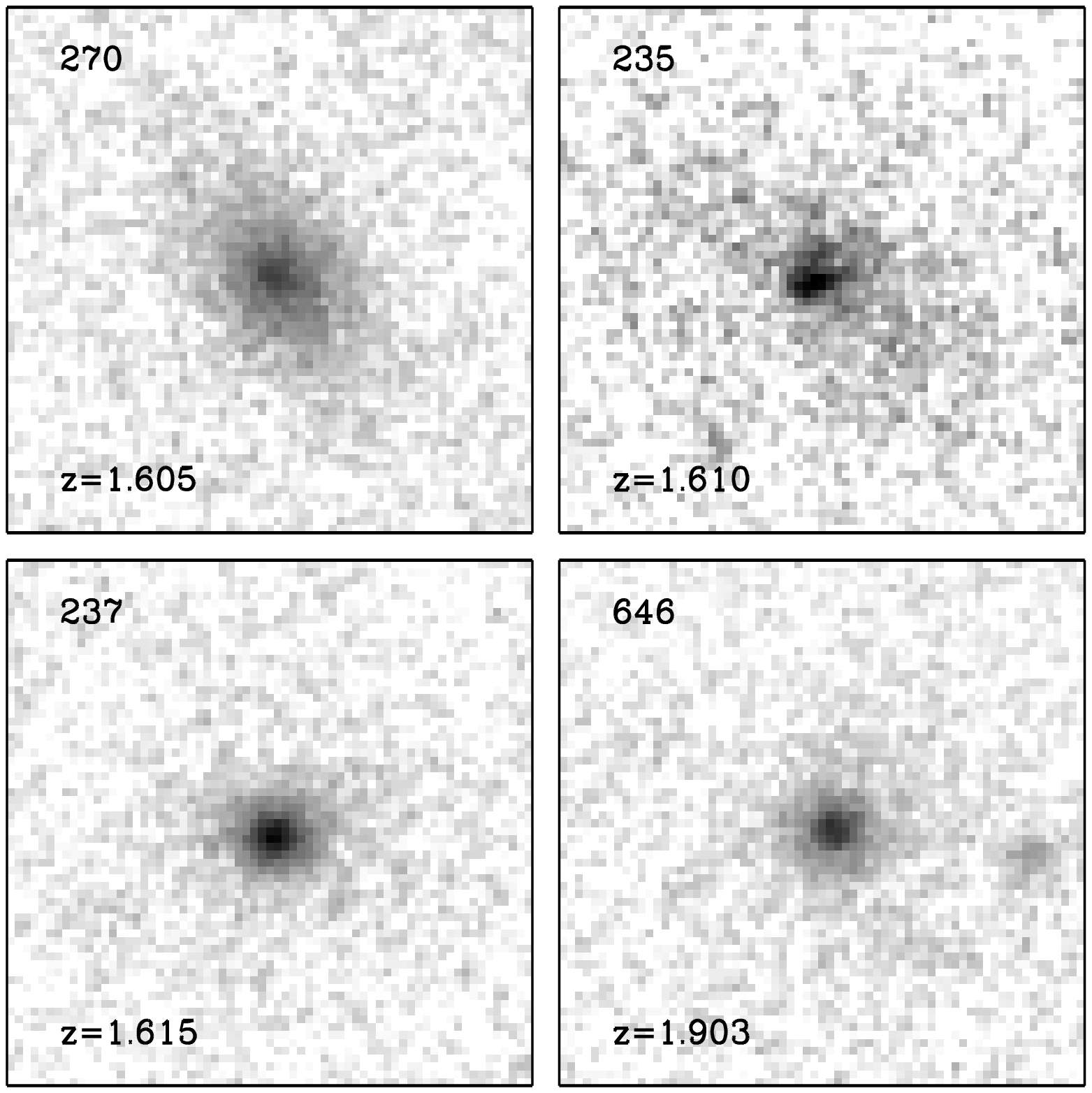}
\caption{}
\label{fig4}
\end{center}
\end{figure}

\pagebreak

\begin{center}
{\bf SUPPLEMENTARY TABLE 1}
\\
\bigskip
{\bf The detected galaxies}
\\
\footnotesize
\bigskip
\begin{tabular}{cccccccl} 
\hline\\
IAU & K20 & R.A. (J2000) & Dec (J2000) & $K_s$ & $R-K_s$ & $z$ & Spectrum\\
 ID  & ID   & h m s & $\circ$ $\prime$ $\prime\prime$& & & & \\
\hline\\
J033210.79-274627.8 & 235 & 03 32 10.776 & -27 46 27.73 & 17.98$\pm$0.04 & 6.47$\pm$0.10&1.610&K20\\
J033210.52-274628.9 & 237 & 03 32 10.507 & -27 46 28.84 & 19.05$\pm$0.05 & 6.83$\pm$0.28&1.615&K20\\
J033212.53-274629.2 & 270 & 03 32 12.525 & -27 46 29.16 & 18.74$\pm$0.05 & 5.99$\pm$0.10&1.605&K20\\
J033233.85-274600.2 & 646 & 03 32 33.847 & -27 46 00.24 & 19.07$\pm$0.07 & 5.99$\pm$0.10&1.903&K20+GOODS\\
\\
\hline
\end{tabular}
\end{center}

\np
\noindent
{\bf Supplementary Table 1}\\ \\
IAU ID: official identification number in the GOODS--South catalog (z-band)\\
(http://www.stsci.edu/science/goods/catalogs).\\ \\
K20 ID: identification number in the K20 survey catalog 
(http://www.arcetri.astro.it/$\sim$k20/).\\ \\
R.A., Dec: Right Ascension and Declination at equinox J2000 based on 
the public ESO/GOODS $K_s$-band VLT+ISAAC image.\\ \\
$K_s$: K20 survey total magnitude in the $K_s$-band (Vega scale).\\ \\
$R-K_s$ color (Vega scale) in 2$^{\prime\prime}$ diameter aperture.\\ \\
$z$: spectroscopic redshift.\\ \\
Spectrum: K20: K20 survey, GOODS: public ESO/GOODS VLT+FORS2 
spectroscopy (Vanzella et al., in preparation; 
http://www.eso.org/science/goods). 

\end{document}